\documentstyle[psfig,amssymb,graphicx,epsf,latexsym,longtable]{mn}
\begin{document}

\title[Optical and Infrared photometry of the type IIn SN 1998S I:
Days 11-146]
{Optical and Infrared photometry of the type IIn SN 1998S:
Days 11-146}
\author[A.Fassia {\rm et al.}]
{A. Fassia$^{1}$\thanks{a.fassia@ic.ac.uk}, W.P.S. Meikle$^{1}$,  W.D. Vacca$^{2}$, S.N. Kemp$^{3,6}$,
N.A. Walton$^{4}$, \and D.L. Pollacco$^{5}$, Smartt S.$^{4}$,  A. Oscoz$^{6}$, 
A. Arag\'{o}n-Salamanca$^{7}$, S. Bennett$^{8}$, 
\and T. G. Hawarden$^{9}$, A. Alonso$^{6}$, D. Alcalde$^{6}$,
A. Pedrosa$^{10}$, J. Telting$^{11}$, 
\and M.J. Arevalo$^{6}$, H.J. Deeg$^{6}$,F. Garz\'{o}n$^{6}$,
A. G\'{o}mez-Rold\'{a}n$^{6}$, G. G\'{o}mez$^{6}$,  
\and C. Guti\'{e}rrez$^{6}$, S. L\'{o}pez$^{6}$, M. Rozas$^{6}$,
M. Serra-Ricart$^{6}$, M.R. Zapatero-Osorio$^{6}$ 
\\ \\
$^1$Astrophysics Group, Blackett Laboratory, Imperial College, Prince
Consort Rd, London SW7 2BZ \\
$^2$Institute for Astronomy, 2680 Woodlawn Dr., Honolulu, HI 96822, USA\\
$^3$Instituto de Astronomia y Meteorologia, Av. Vallarta 2602,
Col. Arcos Vallarta, CP 44130, Guadalajara, Jalisco, Mexico.\\
$^4$Royal Greenwich Observatory, Apartado de Correos 321, 38780 Santa
Cruz de La Palma, Tenerife, Islas Canarias, Spain\\
$^5$Astrophysics and Planetary Sciences Division, The Queen's University
of Belfast, Belfast BT7 1NN \\
$^6$Instituto de Astrofisica de Canarias, C/ Via Lactea S/N, E-38200
La Laguna,Tenerife, Spain\\
$^7$School of Physics \& Astronomy, University of Nottingham, University Park, 
Nottingham NG7 2RD \\
$^8$Institute of Astronomy, Madingley Road, Cambridge CB3 0HA \\
$^9$Joint Astronomy Centre, 660 N. A'Ohoku Place, University Park,
Hilo, Hawaii 96720, USA \\
$^{10}$Centro de Astrofisica da Universidade do Porto, Portugal\\
$^{11}$Isaac Newton Group of Telescopes,NWO (Netherlands Organisation for 
Scientific Research), Apartado 321, \\38700 Santa Cruz de La Palma, Spain\\
}

\maketitle  

\begin{abstract} 
We present contemporaneous optical and infrared photometric
observations of the type~IIn SN~1998S covering the period between 11
and 146 days after discovery.  The infrared data constitute the first
ever infrared light curves of a type IIn supernova.  We use blackbody
and spline fits to the photometry to examine the luminosity evolution.
During the first 2--3 months, the luminosity is dominated by the 
release of shock-deposited energy in the ejecta. After $\sim$100~days
the luminosity is powered mostly by the deposition of radioactive
decay energy from 0.15$\pm$0.05~M$_{\odot}$ of $^{56}$Ni which was
produced in the explosion.  We also report the discovery of an
astonishingly high infrared (IR) excess, $K-L'=2.5$, that was present
at day~130.  We interpret this as being due to thermal emission from
dust grains in the vicinity of the supernova.  We argue that to
produce such a high IR luminosity so soon after the explosion, the
dust must be {\it pre-existing} and so is located in the circumstellar
medium of the progenitor.  The dust could be heated either by the
UV/optical flash (IR echo) or by the X-rays from the interaction of
the ejecta with the circumstellar material.
\end{abstract}


\section{Introduction}
During the past decade a new, distinct subclass of
type II supernovae has been recognised - supernovae type IIn (Schlegel 1990).
In these events the broad absorption components of all lines are weak
or absent (Filippenko 1997).  Their early time spectra are dominated
by strong narrow emission lines.  The appearance of these narrow lines
is attributed to the dynamical interaction of the ejected envelope
with a dense circumstellar wind emitted by the progenitor ({\it e.g.} 
Chugai \& Danziger 1994).  Progenitor mass-loss rates as high as
10$^{-3}$ M$_\odot$ year$^{-1}$ have been estimated from the strength
of these lines (Chugai 1997).  The mass loss can take place in several
episodes sometimes lasting up to the time of explosion.  Consequently,
type~IIn supernovae can provide unique information about the
properties of circumstellar material (CSM) and the late stages of
stellar evolution. In addition, the study of the interaction of these
supernovae with the CSM can provide vital clues about galaxy evolution
and the nature of active galactic nuclei ({\it c.f.} Terlevich {\it et
al.}  1992).

SN~1998S is the brightest type~IIn event so far.  It was discovered on 1998
March 2.68 UT in NGC 3877 by Z. Wan (Li \& Wan 1998) at a broadband
(unfiltered) optical magnitude of +15.2.  By March 18.4 it had brightened
to $V=+12.2$ (see section~2). There was no evidence of the supernova, down
to a limiting apparent magnitude of $\sim$+18, in a prediscovery frame
obtained on 1998 February 23.7 (Leonard {\it et al.} 2000, IAUC 6835).  We
can thus assume that SN~1998S was discovered within a few days of the shock
breakout.  Optical spectra obtained on March~5 by Filippenko \& Moran
(1998) and Huchra (Garnavich {\it et al.} 1998) show prominent H and He
emission lines with narrow peaks and broad wings superimposed on a blue
continuum.  Gerardy {\it et al.} (2000) present near-IR spectra of SN~1998S
spanning 95--355~days post-maximum light.  They identify emission from
carbon monoxide and dust and discuss the environments of these materials.
They also suggest that late-time multi-peak H and He line profiles in their
spectra indicate emission from a disk-shaped or ring-shaped circumstellar
component.  Leonard {\it et al.} (2000) present optical spectroscopy and
spectropolarimetry of SN~1998S spanning the first 494~days after discovery.
They deduce the presence of a dense, asymmetrically distributed CSM
exterior to the expanding ejecta.

Following the discovery of SN~1998S we began a programme of infrared and
optical photometry and spectroscopy.  In this paper we present the
photometry covering the period 11--146~days after discovery.  The spectra
will be described in a separate paper (Fassia {\it et al.}  2000a). The
photometric observations are described in section~2.  In section~3 we
determine the extinction to the supernova. In section~4 we apply blackbody
and spline fits to photometry in order to examine the luminosity evolution
and estimate the ejected $^{56}$Ni mass.  We also consider the observed
high infrared excess.  The work is summarised in section~5.

\begin{figure*}
\vspace{14.0cm}
\includegraphics{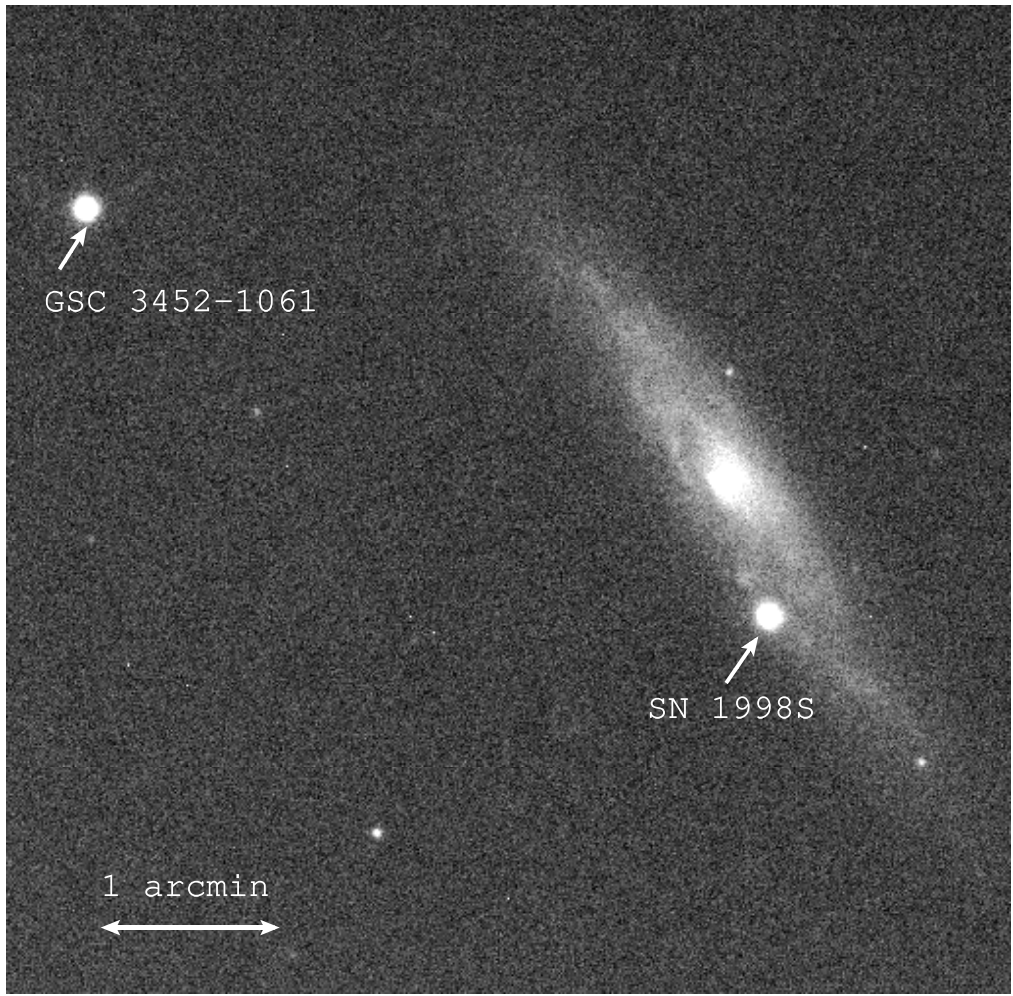}
\caption[]{$R$~band image of SN~1998S in NGC~3877 obtained on 1998 March
16.1~UT at the Jacobus Kapteyn Telescope (JKT), La Palma (north is up and
east is to the left)}
\label{98S}
\end{figure*}

\section{Observations}
SN 1998S is located 16$^{\prime\prime}$ west and 46$^{\prime\prime}$
south of the nucleus of NGC 3877 (Figure~\ref{98S}).  As mentioned
above, SN 1998S was discovered within a few days of the shock
breakout.  We therefore adopt the time of discovery, 1998 March 2.68
UT = JD 2450875.2, as epoch 0~days.
 
\subsection{Optical Photometry}
$BVRI$ images of SN~1998S were obtained during the period from 11 to
146 days. These data were acquired from a variety of telescopes: the
82~cm Instituto de Astrofisica de Canarias telescope (IAC80) on
Tenerife, the 1.0~m Jacobus Kapteyn Telescope (JKT) on La Palma and
the 3.5~m Wisconsin-Indiana-Yale-NOAO telescope (WIYN) at Kitt Peak.  A
list of the observations is given in Table~1.  The IRAF CCDPROC
package was used at Imperial College to reduce the JKT and WIYN data,
following standard procedures.  The IAC80 data were reduced at the IAC
in a similar manner.

Unfortunately, most of the observations were taken under non-photometric
conditions. For these nights, therefore, we performed differential
photometry using for comparison the star GSC 3452-1061 which is located
230.54$^{\prime\prime}$ east and 129.6$^{\prime\prime}$ north of the
supernova (see Figure~1).  Instrumental magnitudes for SN~1998S and the
comparison star were obtained using the aperture photometry package APPHOT
of IRAF.  Since seeing and pixel scales varied between epochs and sites,
the software aperture size for a given image was allowed to vary. Of
course, for each frame the same aperture sizes were used for SN~1998S and
the comparison star.  To calibrate the comparison star, Landolt (1992)
standard fields were observed in $BVRI$ under photometric conditions on
April~11 and 13 at the IAC80, covering a range of airmasses.  Thus, zero
points and transformation coefficients linear in airmass were determined in
$BVRI$.  For the two photometric nights the average magnitudes for GSC
3452-1061 are: $B=+13.06\pm0.03$, $V=+12.570\pm0.007$, $R=+12.280\pm0.008$
and $I=+11.974\pm0.010$.

\begin{figure*}
\vspace{10.0cm}
\includegraphics{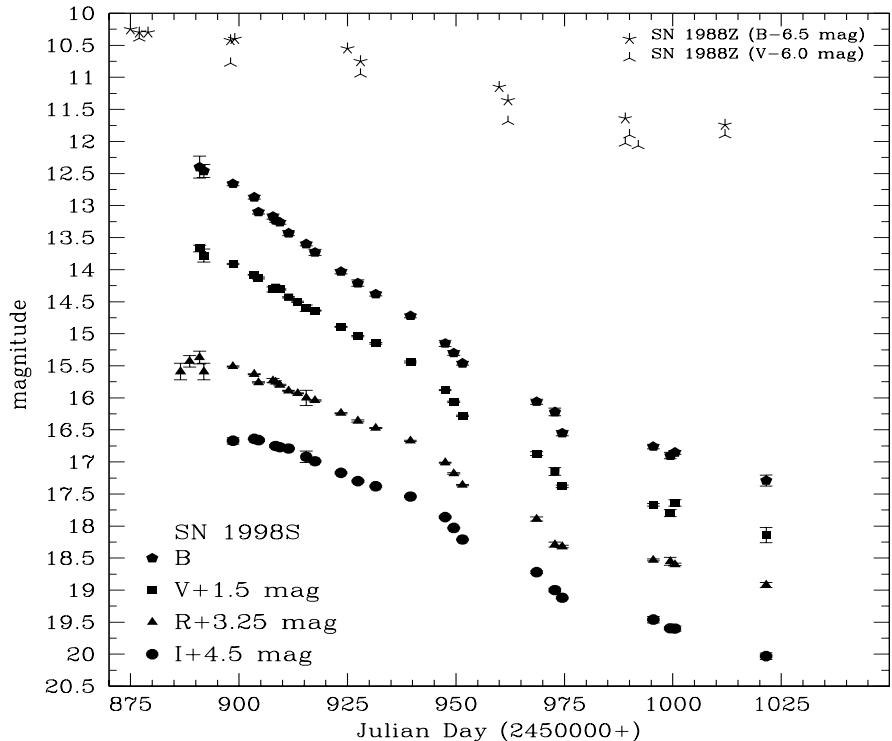}
\caption[]{$BVRI$ light curves of SN~1998S. For comparison we also
show the $B$ and $V$ light curves of the type~IIn SN~1988Z (Turatto
{\it et al.}, 1993)}
\label{98Sbvri}
\end{figure*}

To measure the flux from SN~1998S we had to take into account the
background due to the spiral arm of NGC~3877.  We estimated the background
by examining the galaxy flux in both annular and circular apertures, with a
range of radii, close to the supernova. The annular apertures were centered
on the supernova. The circular apertures were placed at a number of
positions near to the supernova.  The galaxy background determined using
the two types of aperture agreed to within 5\%.  We therefore took the mean
value as the background underlying the supernova and this was used to
correct the instrumental magnitudes.  The final optical photometry for
SN~1998S is listed in Table~\ref{opphot}. The $BVRI$ light curves are shown
in Figure~\ref{98Sbvri}. These light curves agree very well with the $BVRI$
light curves measured by the CFA group (Garnavich {\it et al.} 1999).  For
comparison we also show the $B$ and $V$ light curves of the type~IIn
SN~1988Z (Turatto {\it et al.}, 1993).

\begin{table*}
\centering
\caption[]{Optical photometry of SN1998S}
\begin{minipage}{\linewidth}
\begin{tabular}{llllllll} \hline
JD  & Epoch  & \hspace{0.5cm}$B$ & \hspace{0.5cm}$V$ & \hspace{0.5cm}$R$ & \hspace{0.5cm}$I$    &  \hspace{-0.2cm}Telescope & \hspace{1cm}Observer    \\ 
(2450000+) & (days) &    &    &     &     &    &     \\ 
\hline
886.5 & 11.3  &  -          &   -  &  12.34(0.13) &  -    &   JKT  & A. Aragon-Salamanca, S. Bennett \\
888.6 & 13.4  &  -          &   -  &  12.18(0.09) &  -    &   JKT  & A. Aragon-Salamanca, S. Bennett \\
890.9 & 15.7  & 12.40(0.15) & 12.17(0.05) & 12.12(0.10)  & -       & WIYN & A. Saha \\
891.9 & 16.7  & 12.46(0.1)  & 12.28(0.1)  & 12.34(0.13)  & -       & WIYN & P. Smith \\
898.6 & 23.4  & 12.66(0.03) & 12.41(0.01) & 12.26(0.008) & 12.17(0.05) &
JKT & S. N. Kemp \\
903.5 & 28.3  & 12.87(0.03) & 12.58(0.007)& 12.38(0.008) & 12.14(0.01)
&IAC80 & A. Alonso \\
904.5 & 29.3  & 13.10(0.03) & 12.63(0.015)& 12.41(0.01)  & 12.16(0.01)
&IAC80 & S. N. Kemp, S.L\'{o}pez \\
907.8 & 32.6  & 13.17(0.04) & 12.81(0.04) & 12.48(0.03)  & -          & WIYN & D. Harmer \\ 
908.4 & 33.2  & 13.23(0.03) & 12.78(0.015)& 12.50(0.015) &
12.25(0.015)&IAC80 & H.J.Deeg, S. N. Kemp \\
909.4 & 34.2  & 13.26(0.03) & 12.81(0.01) & 12.55(0.01)  & 12.27(0.01)
&IAC80 & S. N. Kemp \\
911.4 & 36.2  & 13.43(0.04) & 12.93(0.01) & 12.64(0.008) & 12.29(0.01) &IAC80 &  H.J.Deeg \\
913.5  & 38.3  &     -     & 13.00(0.01) & 12.68(0.007) & - & IAC80 &  M.Rozas, F.Garz\'{o}n, A.G\'{o}mez-Rold\'{a}n \\ 
915.5  & 40.3  & 13.60(0.03) & 13.10(0.05) & 12.75(0.12)  & 12.42(0.09) &IAC80 &  M.Rozas \\
917.5  & 42.3  & 13.73(0.045)& 13.14(0.007)& 12.79(0.01)  & 12.49(0.01) &IAC80 &  M.R.Zapatero Osorio\\
923.5  & 48.3  & 14.03(0.035)& 13.39(0.007)& 12.99(0.008) & 12.67(0.01) &IAC80 &  A.Oscoz \\
927.4  & 52.2  & 14.21(0.05) & 13.54(0.01) & 13.11(0.016) & 12.80(0.012)&IAC80 & G.G\'{o}mez, M.J.Arevalo \\
931.5  & 56.3  & 14.38(0.03) & 13.65(0.012)& 13.22(0.008) & 12.88(0.011)&IAC80 & A. Oscoz \\
939.5  & 64.3  & 14.72(0.03) & 13.94(0.014)& 13.42(0.01)  & 13.04(0.01) &JKT  &  A. Pedrosa \\
947.5  & 72.3  & 15.15(0.047)& 14.38(0.008)& 13.76(0.008) & 13.36(0.008)&IAC80 & A. Oscoz \\
949.5  & 74.3  & 15.30(0.038)& 14.56(0.011)& 13.93(0.01)  & 13.53(0.01) &IAC80 & A. Oscoz \\
951.5  & 76.3  & 15.46(0.03) & 14.78(0.01) & 14.11(0.01)  & 13.71(0.01) &IAC80 & A. Oscoz \\
968.6  & 93.4  & 16.06(0.04) & 15.37(0.03) & 14.64(0.03)  & 14.22(0.02) &
WIYN/JKT & D. Harmer/ S. N. Kemp \\ 
972.8  & 97.6  & 16.22(0.06) & 15.65(0.06) & 15.04(0.04)  & 14.50(0.03) & WIYN & D. Harmer \\  
974.5  & 99.3  & 16.55(0.042)& 15.88(0.02) & 15.07(0.02)  & 14.62(0.013)&JKT   & J. Telting \\
995.5  & 120.3 & 16.76(0.035)& 16.17(0.02) & 15.28(0.015) & 14.96(0.05) &IAC80 & A.Oscoz, D.Alcalde \\
999.4  & 124.2 & 16.90(0.05) & 16.30(0.055)& 15.30(0.06)  & 15.096(0.025)& IAC80 & A. Oscoz \\
1000.5 & 125.3 & 16.85(0.03) & 16.14(0.05) & 15.35(0.02)  & 15.10(0.04)
&IAC80 & S. N. Kemp \\
1021.5 & 146.3 & 17.29(0.085)& 16.64(0.12)  & 15.67(0.04)& 15.53(0.05) & IAC80 & A. Alonso, C. Guti\'{e}rrez\\
\hline
\end{tabular} \\
NOTE.- Figures in brackets give the internal error of each
magnitude
\end{minipage}
\label{opphot}
\end{table*}
\begin{table*}
\centering
\caption[]{Infrared photometry of SN1998S}
\begin{minipage}{\linewidth}
\begin{tabular}{llllllll} \hline
JD  & Epoch  & \hspace{0.5cm}$J$ & \hspace{0.5cm}$H$ & \hspace{0.5cm}$K$ & \hspace{0.5cm}$L$    &  \hspace{-0.2cm}Telescope & \hspace{0.2cm}Observer    \\ 
(2450000+) & (days) &    &    &     &     &    &     \\ 
\hline
889.0  & 13.8  & 12.11(0.01) & 12.06(0.01)& 12.05(0.02) &            & IRTF   & Bill Vacca \\
890.0  & 14.8  & 12.07(0.01) & 12.02(0.01)& 11.98(0.02) &            & IRTF   & Bill Vacca \\
890.9  & 15.7  & 12.06(0.01) & 12.01(0.01)& 11.98(0.02) &            & IRTF   & Bill Vacca \\
891.9  & 16.7  & 12.02(0.01) & 11.97(0.01)& 11.93(0.04) &            & IRTF   & Bill Vacca \\
892.9  & 17.7  & 12.02(0.01) & 11.93(0.02)& 11.90(0.02) &            & IRTF   & Bill Vacca \\
924.9  & 49.5  & 12.37(0.01) &  -           & -             &        & IRTF   & Bill Vacca \\
951.9  & 76.7  & 13.36(0.02) & 13.05(0.03)  & 12.70(0.04) &            & IRTF   & Bill Vacca \\
976.8  & 101.6 & 14.25(0.02) & 13.92(0.02)  & 13.31(0.05)   &  -     & UH2.2m    & Bill Vacca \\
984.8  & 109.6 & 14.33(0.01) & 14.02(0.02)  & 13.43(0.11) &            & IRTF   & Bill Vacca \\
1003.9 & 128.7 & 14.73(0.06) & 14.34(0.06)  & 13.55(0.01)   &          - & UKIRT &Tim Hawarden \\
1004.9 & 129.7 &    -          &       -      & 13.55(0.01)  & 11.0(0.15) & UKIRT &Tim Hawarden  \\  
\hline
\end{tabular} \\
NOTE.- Figures in brackets give the internal error of each 
magnitude
\end{minipage}
\label{irphot}
\end{table*}

\begin{figure*}
\vspace{8.7cm}
\includegraphics{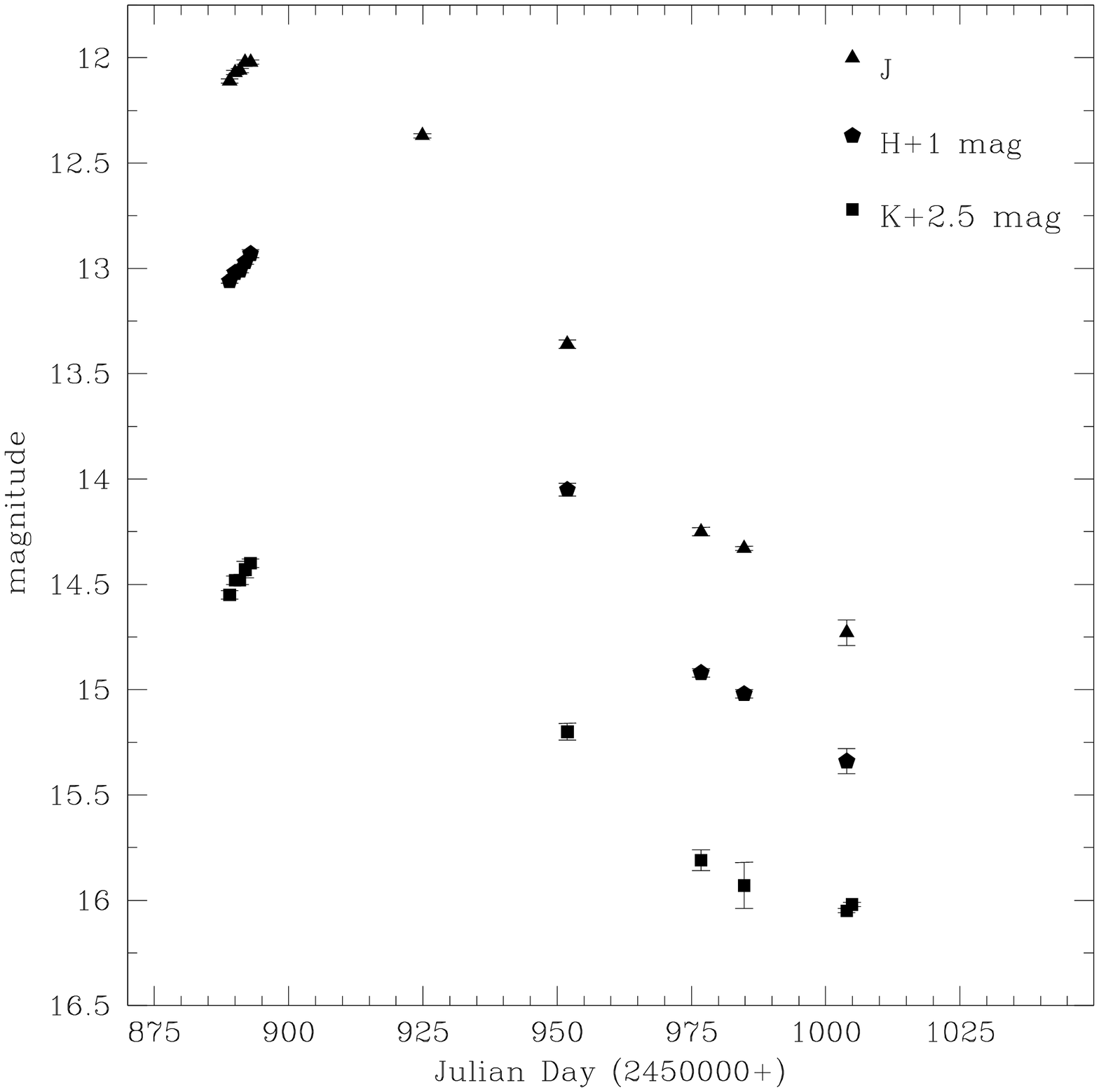}
\caption[]{$JHK$ light curves of SN 1998S}
\label{98Sjhk}
\end{figure*}

\subsection{Infrared Photometry} 
$JHK$ broadband images of SN 1998S were obtained in the period 14 to
130 days. These data were acquired at the Infrared Telescope Facility
(IRTF), the University of Hawaii Telescope (UH2.2m) and the United Kingdom
Infrared Telescope (UKIRT), all on Mauna Kea.  In addition, an $L'$ image
was obtained at UKIRT on day~130. The list of these observations is
given in Table~\ref{irphot}.  All the infrared observations reported
here were obtained under photometric conditions.

The IRTF data were acquired with the NSFCam infrared camera, a 1-5
$\mu$m imager which incorporates a 256$\times$256 InSb array. For these
observations, a plate scale of 0.3$^{\prime\prime}$/pixel was used;
this yields a field of view of 76.8$^{\prime\prime}$. Images were
obtained in the standard $J$, $H$, and $K$ broadband filters. In each
filter, images of the supernova were acquired in a 7-point dither
pattern, interspersed with equal integrations on an offset sky field
which contained a bright star.  Occasionally, two sets of dithered
observations were obtained for a given filter.  Each image consists of
5-10 coadded frames. Integration times for each frame were on the order
of a few seconds.  Images of UKIRT Faint Standard Stars FS21 and FS23
were obtained immediately before and after the SN images.

The UKIRT data were obtained using the IRCAM3 camera. This contains a
256$\times$256 InSb array with a plate scale of
0.281$^{\prime\prime}$/pixel. For the observations a 5~point dither pattern
and an integration time of few seconds was adopted.  The UH2.2m data were
acquired with the Quick Infrared Camera (QUIRC), which incorporates a
1024$\times$1024 HgCdTe array (Hodapp {\it et al.} 1996). The broadband filters
$J$, $H$, and $K'$ and an integration time of 30 s were used for the
observations. The plate scale was 0.19$^{\prime\prime}$/pixel.

The IRTF data were reduced using IRAF scripts in the standard manner. Flat
fields and sky frames were produced from the offset field images. Aperture
photometry was carried out on each flat-fielded and sky-subtracted image
using APPHOT in IRAF. An aperture radius of 12 pixels
(3.6$^{\prime\prime}$) was used; the background was estimated from a
5-pixel-wide annulus with an inner radius of 15 pixels from the center of
each object.  Instrumental magnitudes were converted to standard magnitudes
using the observations of the UKIRT Faint Standard Stars and the published
magnitudes given by Hunt {\it et al.}  (1998).  Because the standard stars
were observed over a very limited range in airmass and have a limited range
in color, we adopted the median extinction coefficients for Mauna Kea in
each passband, as given by Krisciunas {\it et al.} (1987), and the average
color term derived from both these and previous observations of other SNe
with the same observational set up. Therefore, we fitted only the zero
points, and applied the zero points and the color and extinction
coefficients to the instrumental mags of the SN and the star in the offset
field. Uncertainties were derived from the standard deviations of the 7-14
instrumental magnitude estimates of the SN in each filter, which were then
carried through the photometric transformation equations. The errors
presented in Table 2 are statistical only; the observations of the star in
the offset field provide a measure of the systematic errors, which are less
than about 0.04 mags.

The observations and data reduction of the UH2.2m data were carried
out in a manner similar to that employed for the IRTF/NSFCam images.  A
source aperture with a radius of 15 pixels (2.8$^{\prime\prime}$) was
used for the photometry. Flux calibration was carried out using images
of UKIRT Standards FS21, FS23, FS27, and FS35.  Because multiple
observations of these stars were acquired, a full solution for the
photometric coefficients (zero point, color coefficient, and extinction
coefficient) could be obtained.  $K$-band spectra taken by Fassia {\it
et al.} (2000a) and Gerardy {\it et al.} (2000) on days 109 and 110
respectively show strong emission from the first overtone of the CO
molecule. To take into account the effect of the CO emission the
transformation of the $K'$ mags to $K$ mags was performed as
follows. We created transmission functions for the $K'$ and $K$ bands
using the UH2.2m filter functions and the standard Mauna Kea
atmospheric transmission.  We then multiplied the $K$-band spectrum by
the $K'$ band transmission function and hence derived an apparent
magnitude. Assuming that the CO emission remained constant between
101.6d (UH2.2m photometry) and 109d ($K$-band spectra) we then scaled
the $K$-band spectrum to obtain a match of the apparent magnitude to
the $K'$ photometry. We then multiplied the {\it scaled} $K$-band
spectrum with the $K$-band transmission function and determined the $K$
magnitude of SN 1998S on day 101.6.

The UKIRT data were reduced using the Starlink package IRCAMDR (Aspin
1996). They were first corrected for the dark-current.  Median flatfields
created from the target images were used for the flatfield correction. The
resulting frames were then mosaiced together.  Instrumental magnitudes were
measured using the aperture photometry package APPHOT in IRAF.  The
determination of the galaxy background was performed in the same way as for
the optical data ({\it cf.} Section 2.1).  Calibration of the $JHK$ data was
relative to the standard star FS21 (airmass corrected).  Calibration of the
$L'$ image was relative to HD105601 ($L'=+6.65$) and HD106965 ($L'=+7.34$).

The $JHK$ magnitudes are listed in Table~\ref{irphot}.  The $JHK$ data
are plotted in Figure~\ref{98Sjhk}.  To our knowledge these represent
the first ever determination of the infrared light curves of a type
IIn supernova. We note the astonishingly large infrared excess of
$K-L'=2.55\pm0.15$ exhibited at just 130~days.  This excess persisted
for more than a year and a half after discovery (Fassia {\it et al.}
2000b) and it is discussed in Section 4.2.

\subsection{The colours} 
\begin{figure}
\vspace{16cm}
\includegraphics{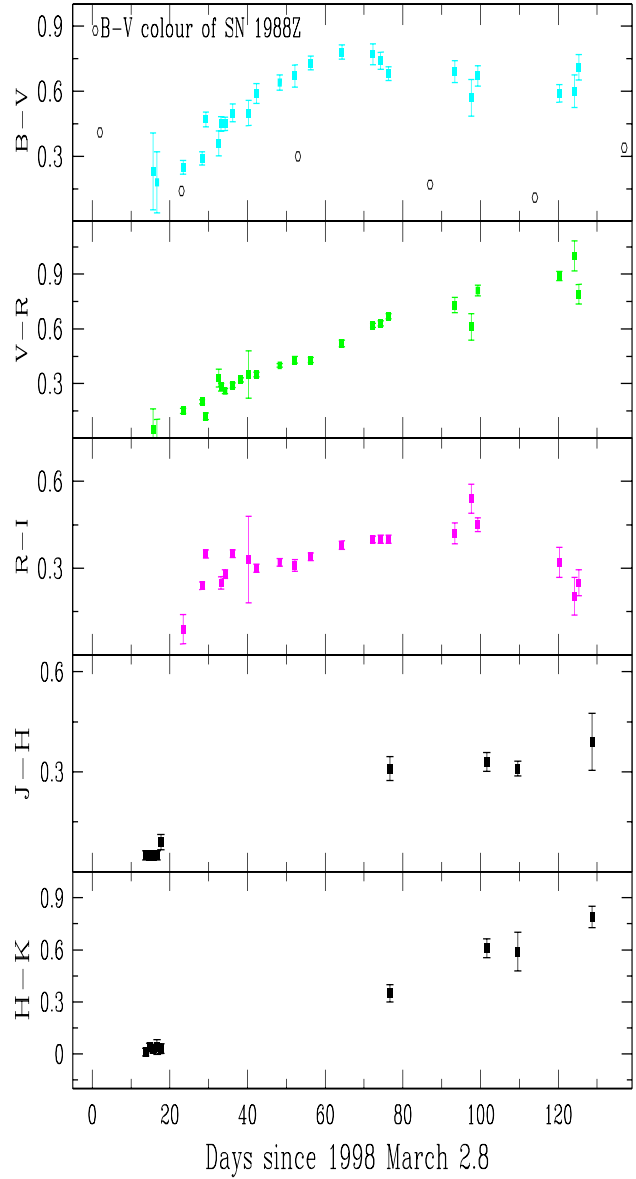}
\caption[]{Evolution of $B-V$, $V-R$, $R-I$, $J-H$ and $H-K$ colours of
SN~1998S. Also shown for comparison is the $B-V$ colour evolution of
SN~1998Z (Turatto {\it et al.}, 1993).}
\label{98Soplc}
\end{figure}

In Fig.~\ref{98Soplc} we present the optical and IR
colour evolution of SN 1998S.  In general, reddening of the colours took
place between days~15 and ~70.  This is consistent with photospheric
cooling, and is discussed in section~4.1 ({\it cf.} Fig~\ref{lum_temp}.)
Between days 80 and 130 the various colour indices evolve differently.  The
$R$ and $I$ bands are significantly affected by the evolution of the
H$\alpha$ and Ca~II triplet lines (Fassia {\it et al.} 2000a) .  The
increasing prominence of H$\alpha$ was responsible for the continued
reddening in $V-R$, while the strength of the Ca~II triplet relative to
H$\alpha$ accounts for much of the $R-I$ evolution.  In the IR, the $H-K$
colour continued to redden right up to the last observation on day~130.
This is discussed in Section~4.3.

\section{Extinction}
\begin{figure}
\vspace{9cm}
\includegraphics{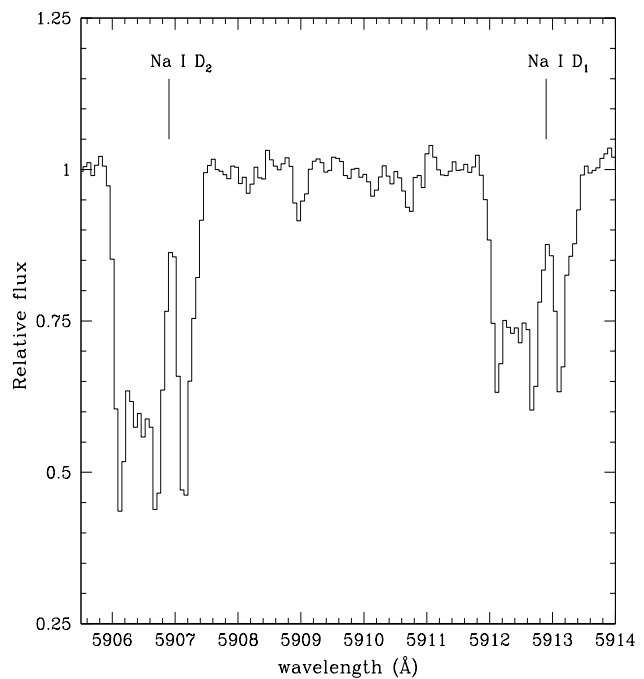}
\caption[]{High-resolution spectra of interstellar Na~I~D absorption lines 
towards SN~1998S.  These were obtained at the William Herschel
Telescope, La Palma on 1998, April 8.0} 
\label{fignaD}
\end{figure}

In order to determine the extinction towards SN~1998S we examined
interstellar Na~I~D lines in high-resolution spectrum obtained at the
William Herschel Telescope, La Palma, on 1998 April 8.0 (Bowen {\it et
al.} 2000; Fassia {\it et al.} 2000a).  These are shown in
Fig.~\ref{fignaD}.  The spectrum is of sufficiently high signal-to-noise
ratio and resolution (7.4 km/s FWHM) to allow us to determine the physical
parameters of the interstellar absorbing clouds.  We fitted theoretical
Voigt profiles to the data (see Carswell {\it et al.} 1991 for details).
For each feature in the spectra a model profile was computed for an initial
set of values of the column density, $N(Na~I)$, the heliocentric velocity,
$v_{H}$, and the Doppler parameter ({\it i.e.} the velocity dispersion
within a cloud), $b$$=\sqrt{2} \sigma$.  The instrumental profile was
approximated by a Gaussian whose FWHM is determined from fits to unblended
emission lines from the comparison lamp.  The theoretical profiles were
then convolved with the instrumental response function.  Finally, the model
parameters were adjusted to match the observed profiles by minimising
$\chi^{2}$.

Several constraints were imposed in the profile fitting procedure.
For each feature we required that both components of the doublet were
fitted.  The number of absorption systems ({\it i.e.} clouds) was kept
to a minimum by demanding that only features which were at least
partially resolved were included in the fit.  In the cases where a
large range of values were obtained for $b$, we chose the largest
value and then repeated the fit with only $N(Na~I)$ and $v_{H}$ as the
free parameters.  For such cases this approach yields a lower limit
to $N(Na~I)$ (Nachman \& Hobbs 1973).

\begin{figure}
\vspace{9.0cm}
\includegraphics{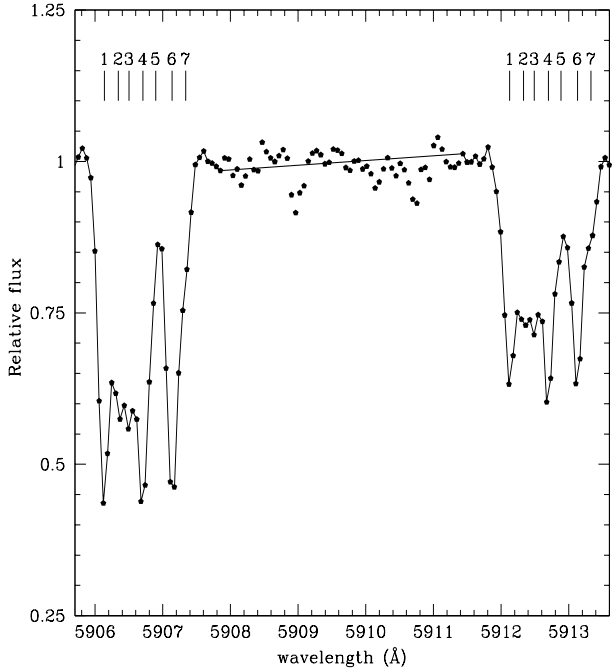}
\caption[]{Comparison of the observed intensities of the intestellar
Na~I~D lines ({\it dots}) and the best-fit model profiles ({\it
continuous lines}) computed as described in Section~3.  Details of the
model fits are given in Table~\ref{voigt}.  Velocities of the
individual interstellar clouds are given in the table, and identified
here by vertical tick marks.}
\label{figvoigt}
\end{figure}

The final fits are shown in Figure~\ref{figvoigt} superimposed on the
observed data points.  The parameters of the model fits are presented in
Table~\ref{voigt} where for each absorption system we give the heliocentric
velocity, $v_{h}$, the Doppler parameter, $b$, and the column density,
$N$(Na~I).  Seven individual velocity systems are identified, with
$v_{H}\sim$+824 to +885~km/s.  Also given are the directly measured
equivalent widths for the blends of systems 1--5 and 6--7.  Measurements of
the systemic velocity of the host galaxy, NGC~3877, vary from +838~km/s
using optical observations of the nuclear region (Ho {\it et al.} 1995), to
+902~km/s, using the H~I 21~cm emission line (De Vaucouleurs {\it et al.}
1991).  Thus, the absorption systems are clearly associated with NGC 3877.
For each absorbing cloud, the derived $N(Na~I)$ can be related to the total
hydrogen column density, $N$(H)=$N$(H~I) + $N$(H~II) by adopting the
correlation between these quantities established for the Milky Way Galaxy.
Hobbs (1974, 1976) showed that $N$(Na~I)$\sim$($N$(H))$^{2}$.  However, for
measurements that sampled lower column-density regions, Ferlet,
Vidal-Madjar \& Gry (1985) showed that the relation between $N$(Na~I) and
$N$(H) is close to linear for $N$(H)$\leq$10$^{21}$cm$^{-2}$ (corresponding
to $N$(Na~I)$\leq$5.6$\times$10$^{12}$cm$^{-2}$ ).  In particular they
found that $logN$(Na I)=(1.04$\pm$0.08)$logN$(H)$-$9.09 with a 1$\sigma$
scatter of $\pm$3dex.  Given the column densities in Table~\ref{voigt}, we
therefore used this relation to derive $N$(H) from our $N$(Na I)
measurements.  The $N$(H) values were then used to find the extinction
towards SN~1998S. In doing this, we assumed the same gas-to-dust ratio in
NGC 3877 as in our Galaxy {\it i.e.} we adopted the relation:
$N$(H)=5.8$\times$10$^{21}E(B-V)$ (Bohlin {\it et al.} 1978).  Summing over
all the absorption systems we get $E(B-V)=0.18^{+0.18}_{-0.09}$.
\begin{table*}
\centering
\caption{Measurements of the Na D interstellar absorption lines }
\begin{minipage}{\linewidth}
\renewcommand{\thefootnote}{\thempfootnote}
\begin{tabular}{ccccccc} \hline
System   & $v_{h}$  & $b$   &  $logN$(Na I) & $logN$(H) &
W$_{\lambda}$(D$_{2}$) & W$_{\lambda}$(D$_{1}$)\\
Number  &  (km/sec) & (km/sec) &  (cm$^{-2}$)  & (cm$^{-2}$) & (m\AA)
&(m\AA) \\
\hline 
 1 & 824.2$\pm$0.2 & 3.68$\pm$0.22  &  11.99$\pm$0.01  &  20.26$\pm$0.3 &
366$\pm$25 \footnote{Sum of systems 1-5, which are blended together}   & 245$\pm$30$^{\footnotesize{a}}$   \\
 2 & 834.7$\pm$0.6 & 1.45$\pm$0.43  &  11.79$\pm$0.07  &  20.07$\pm$0.3  &     &   \\   
 3 & 842.8$\pm$0.6 & 2.39$\pm$0.78  &  11.76$\pm$0.04  &  20.05$\pm$0.3   &     &    \\ 
 4 & 853.3$\pm$0.3 & 3.35$\pm$0.53  &  12.01$\pm$0.01  &  20.28$\pm$0.3&     &    \\ 
 5 & 862.9$\pm$0.9 & 0.30$\pm$0.07  &  12.12$\pm$0.30  &  20.39$\pm$0.3&     &    \\ 
 6 & 875.2$\pm$0.3 & 3.48$\pm$0.35  &  11.99$\pm$0.01  &  20.26$\pm$0.3
&130$\pm$15 \footnote{Sum of systems 6-7, which are blended together}
& 87$\pm$17$^{\footnotesize{b}}$     \\  
 7 & 885.4$\pm$0.3 & 1.47$\pm$1.21  &  11.31$\pm$0.44  &  19.61$\pm$0.3   &     &    \\
\hline 
\end{tabular} 
\end{minipage}
\label{voigt}
\end{table*}

Given the large uncertainty in the extinction derived from the
column density of Na~I we also employed a different approach and
estimated the extinction of SN 1998S using empirical relations between
equivalent widths of the Na~I~D lines and the reddening $E(B-V)$. The
total equivalent widths of the Na~I~D lines summing over all the
components are $W_{\lambda}(D_{1})$=330$\pm$35$m\AA$,
$W_{\lambda}(D_{2})$=500$\pm$30$m\AA$ and
$W_{\lambda}(D_{1,2})$=828$\pm$45$m\AA$.  Barbon {\it et al.} (1990)
provided a sample of supernovae for which both the reddening and the
equivalent width of the Na~I~D lines was known.  A linear fit to these
observations gives $E(B-V)=0.27(\pm0.04) \times
W_{\lambda}(D_{1,2})$. From this we get $E(B-V)=0.22\pm0.04$ for SN
1998S.  Munari \& Zwitter also provided relations between equivalent
widths of the Na~I~D lines and reddening using observations from early
B stars.  Using their results we get $E(B-V)=0.21\pm0.08$ for SN
1998S. Finally, using the linear fits that Richmond {\it et al.} 1994,
applied to the Sembach {\it et al.} (1993) sample of 57 stars in the
Milky Way we get $E(B-V)=0.17$ from the $D_{1}$ lines and $E(B-V)=0.20$
from the $D_{2}$ lines. Averaging these two results and taking into
account the scatter among the points of the Sembach {\it et al.} sample
we estimate $E(B-V)=0.18\pm0.10$. We adopt the mean value of $E(B-V)$
found by all the methods described above (including the $E(B-V)$
estimate derived from the  Na~I column density) and get
$E(B-V)=0.20^{+0.11}_{-0.08}$. However, given that all the
estimates are based on the Na~I~D column density, we recognise that
the values are probably not truly independent.  Moreover, to the
extent that the correlation between Na~I column density and extinction
is uncertain all the $E(B-V)$values may be subject to systematic
error. Adding reddening due to dust in the
Milky Way {\it viz.}  $E(B-V)_{MW}=0.023\pm0.004$ (Schlegel {\it et
al.} 1998) we get a total $E(B-V)=0.22^{+0.11}_{-0.08}$.  Adopting
A$_{V}=3.1E(B-V)$ (Whitford 1958) for NGC~3877, this translates to
$A_{V}=0.68^{+0.34}_{-0.25}$. 

\section{Analysis and Discussion}
\subsection {Total Luminosity}
In order to examine the energetics of SN~1998S, we used the photometry to
examine the total luminosity as follows.  For each date on which we
acquired photometry in {\it any} waveband we assembled a complete set of
$UBVRIJHK$ magnitudes by interpolating to the specific date using both the
data presented here and those published by (Garnavich {\it et al.}  1999).
The $U$-band data were taken exclusively from Garnavich {\it et al}.
Magnitudes were converted to fluxes using the calibrations of Wilson {\it
et al.}  (1972) and Bessel (1979).  Adopting the $V$-band extinction value
derived in the previous section, the fluxes were then de-reddened using the
Cardelli {\it et al.} (1989) law.

\begin{figure}
\vspace{10cm}
\includegraphics{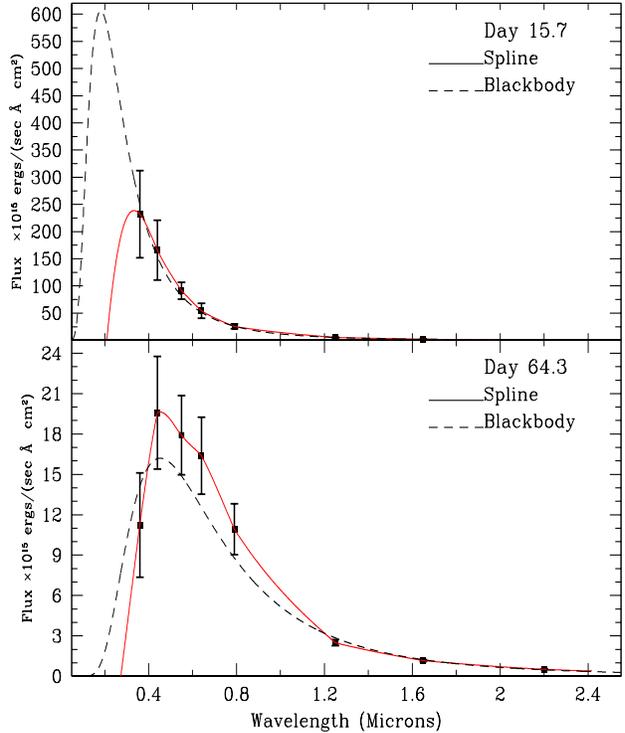}
\caption[]{Sample spline and blackbody fits to the photometry of SN
1998S on days 15.7 and 64.3. }
\label{examp}
\end{figure}

\begin{figure*}
\vspace{15.0cm}
\includegraphics{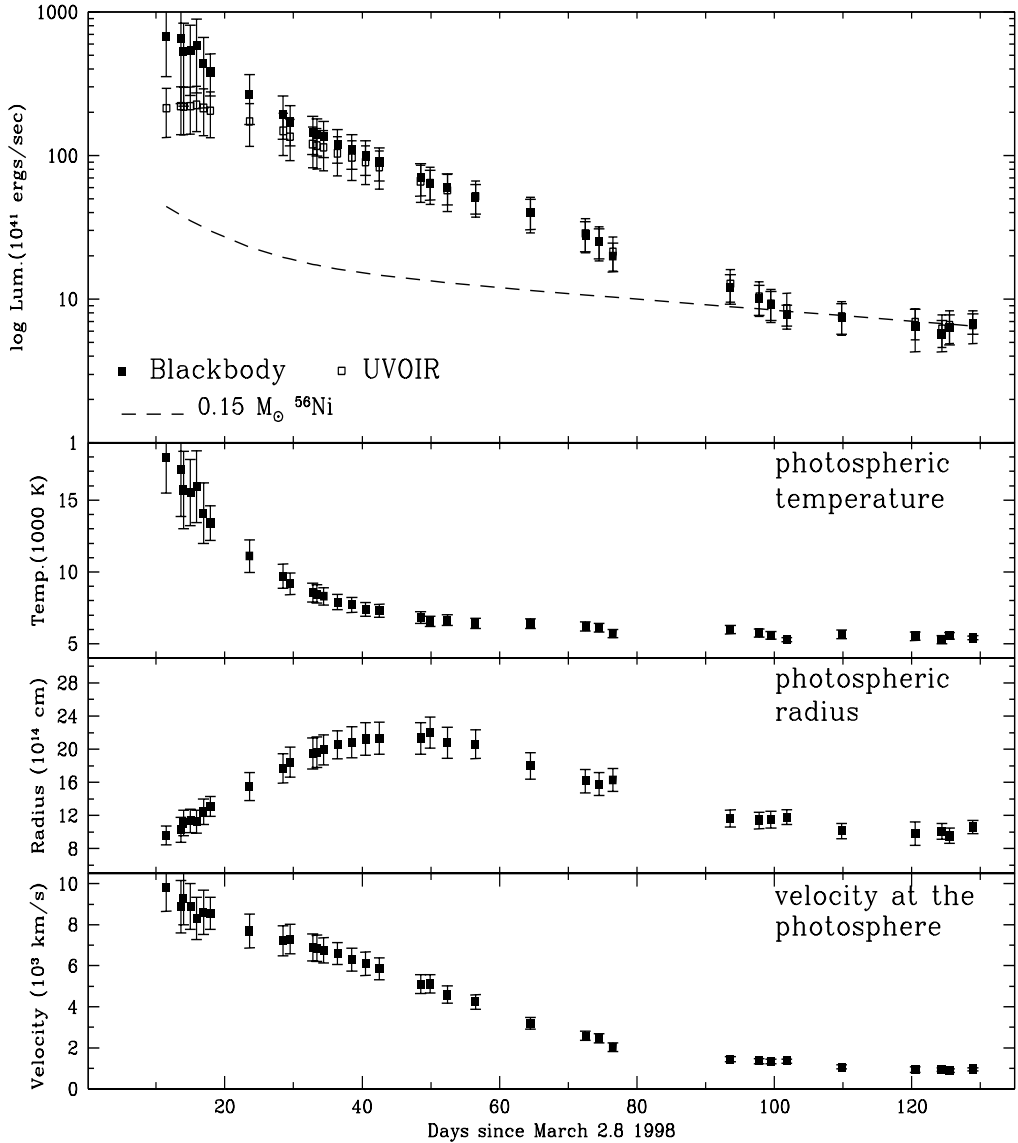}
\caption[]{Upper diagram: Blackbody and $UVOIR$ light curves derived
respectively from blackbody fits (filled squares) and spline fits (open
squares). Also shown (dashed line) is the total luminosity of
0.15~M$_{\odot}$ of $^{56}$Ni. Lower diagram: photospheric temperatures,
radiiand velocities at photsphere derived from blackbody fits}
\label{lum_temp}
\end{figure*}

In one procedure, we fitted blackbody functions to the de-reddened
$UBVRIJHK$ data via $\chi^{2}$ minimisation, and then integrated under the
best-fit blackbody curve to find the total luminosity.  We shall refer to
this as the ``blackbody luminosity''.  As well as providing total fluxes,
this method also provided estimates of the photospheric temperature (T) and
radius (R). We also estimated the velocity of material {\it at} the
photosphere, V$_{ph}$, by dividing R by the corresponding epoch.  In a
second approach we fitted spline curves to the data (Press {\it et al.}
1992), and again integrated under the curve.  However, in this case the
integration range was fixed by the limits of the filter functions used in
the observations {\it viz.}  $\lambda_{start}$=3500 \AA\ and
$\lambda_{finish}$=25000 \AA ({\it i.e. we did not extrapolate beyond this
wavelength range}).  This procedure provided what we shall call the
``$UVOIR$ luminosity''.  As well as avoiding significant extrapolations
into the UV, this method made no assumptions about the exact shape of the
spectrum in the $U$ to $K$ range.  Examples of blackbody and spline fits
are shown for representative days in Fig.~\ref{examp}.  The estimated
fluxes were converted to luminosities by adopting a distance of
17.0$\pm$1.2~Mpc to NGC~3877 (Tully 1988). These are listed in
Table~\ref{lum}, together with the temperatures, radii and velocities given
by the blackbody fits, and are plotted in Fig.~\ref{lum_temp}. Errors in the
above quantities were estimated from the quality of the fits taking also
into account uncertainties in the extinction and distance.

\begin{table*}
\centering
\caption[]{ Luminosities, photospheric temperatures, photospheric
radii and velocity of material at the photosphere for SN1998S} 
\begin{minipage}{\linewidth}
\begin{tabular}{ccccccccc} \hline
JD & Epoch & T  & R\footnote{Radii, velocities and luminosities have been calculated assuming that the
distance of SN 1998S is 17~Mpc (Tully 1988, Nearby Galaxies Catalogue)}
& V$_{ph}$ & F$_{Blackbody}$ &
 F$_{Spline}$ & L$_{Blackbody}$ & L$_{Spline}$ \\ 
(2450000+) & (d) & (K) & (10$^{14}$cm)& (km/s) & (10$^{-10}$ erg  & (10$^{-10}$ erg
& (10$^{41}$ erg & (10$^{41}$ erg\\
&&&&&sec$^{-1}$~cm$^{-2}$) & sec$^{-1}$~cm$^{-2}$) & sec$^{-1}$) & sec$^{-1}$) \\
\hline
  886.5 & 11.3  & 18000(2500) &  9.60(1.12)  & 9800(1143) & 19.67(9.77) &  6.18(2.15) & 680.0(324.5) & 213.6( 79.9) \\
  888.6 & 13.4  & 17150(3300) & 10.27(1.48)  & 8870(1278) & 18.80(12.4) &  6.40(2.17) & 650.0(419.4) & 221.2( 81.0)\\
  889.0 & 13.8  & 15710(2700) & 11.10(1.51)  & 9270(1264) & 15.33(9.18) &  6.35(2.14) & 530.0(308.9) & 219.5( 79.9)\\
  890.0 & 14.8  & 15540(2300) & 11.35(1.43)  & 8900(1119) & 15.48(8.19) &  6.38(2.13) & 535.0(273.2) & 220.5( 79.6)\\
  890.9 & 15.7  & 15950(2500) & 11.26(1.40)  & 8300(1028) & 16.92(9.29) &  6.52(2.10) & 585.0(310.9) & 225.4( 78.9)\\
  891.9 & 16.7  & 14100(2100) & 12.45(1.56)  & 8600(1079) & 12.73(6.75) &  6.20(2.05) & 440.0(225.3) & 214.3( 76.8)\\
  892.9 & 17.7  & 13400(1200) & 13.10(1.21)  & 8560( 788) & 11.14(3.94) &  5.95(1.93) & 385.0(125.4) & 205.7( 72.5)\\
  898.6 & 23.4  & 11100(1150) & 15.50(1.68)  & 7700( 831) &  7.67(3.09) &  5.01(1.50) & 265.0(100.5) & 173.2( 57.1)\\
  903.5 & 28.3  &  9700( 850) & 17.70(1.78)  & 7220( 728) &  5.64(2.02) &  4.30(1.27) & 195.0( 64.6) & 148.6( 48.4)\\
  904.5 & 29.3  &  9170( 750) & 18.44(1.82)  & 7300( 717) &  4.92(1.69) &  3.92(1.14) & 170.0( 53.5) & 135.5( 43.6)\\
  907.8 & 32.6  &  8560( 650) & 19.50(1.87)  & 6900( 663) &  4.19(1.38) &  3.48(1.00) & 145.0( 43.2) & 120.3( 38.3)\\
  908.4 & 33.2  &  8470( 630) & 19.63(1.84)  & 6850( 641) &  4.05(1.31) &  3.40(0.96) & 140.0( 40.8) & 117.5( 36.9)\\
  909.4 & 34.2  &  8300( 600) & 19.90(1.82)  & 6750( 616) &  3.91(1.23) &  3.30(0.92) & 135.0( 38.3) & 114.1( 35.5)\\
  911.4 & 36.2  &  7900( 540) & 20.54(1.68)  & 6600( 535) &  3.47(1.03) &  3.00(0.82) & 120.0( 31.4) & 103.7( 31.7)\\
  913.5 & 38.3  &  7700( 520) & 20.85(1.87)  & 6300( 565) &  3.18(0.97) &  2.80(0.77) & 110.0( 29.8) &  96.8( 29.8)\\
  915.5 & 40.3  &  7400( 480) & 21.24(1.99)  & 6100( 571) &  2.89(0.88) &  2.60(0.70) & 100.0( 27.0) &  89.9( 27.2)\\
  917.5 & 42.3  &  7300( 450) & 21.35(1.93)  & 5850( 527) &  2.60(0.76) &  2.40(0.63) &  90.0( 23.3) &  83.0( 24.6)\\
  923.5 & 48.3  &  6830( 400) & 21.30(1.93)  & 5100( 461) &  2.02(0.58) &  1.92(0.49) &  70.0( 17.7) &  66.4( 19.2)\\
  924.9 & 49.5  &  6560( 350) & 22.00(1.87)  & 5120( 435) &  1.85(0.50) &  1.86(0.47) &  64.0( 15.0) &  64.3( 18.5)\\
  927.4 & 52.2  &  6640( 370) & 20.80(1.87)  & 4600( 414) &  1.74(0.49) &  1.66(0.42) &  60.0( 14.7) &  57.4( 16.5)\\
  931.5 & 56.3  &  6420( 345) & 20.60(1.73)  & 4230( 356) &  1.48(0.40) &  1.50(0.37) &  51.0( 11.9) &  51.9( 14.6)\\
  939.5 & 64.3  &  6400( 340) & 18.00(1.59)  & 3200( 286) &  1.16(0.32) &  1.16(0.28) &  40.0(  9.5) &  40.1( 11.1)\\
  947.5 & 72.3  &  6200( 320) & 16.15(1.42)  & 2580( 226) &  0.81(0.22) &  0.83(0.19) &  28.0(  6.6) &  28.7(  7.7)\\
  949.5 & 74.3  &  6120( 310) & 15.80(1.41)  & 2460( 219) &  0.72(0.20) &  0.73(0.17) &  25.0(  5.9) &  25.2(  6.8)\\
  951.5 & 76.3  &  5700( 280) & 16.30(1.38)  & 2020( 209) &  0.58(0.15) &  0.62(0.14) &  20.0(  4.5) &  21.4(  5.7)\\
  968.6 & 93.4  &  5980( 290) & 11.64(1.03)  & 1440( 127) &  0.35(0.09) &  0.37(0.08) &  12.0(  2.8) &  12.8(  3.3)\\
  972.8 & 97.6  &  5750( 290) & 11.40(1.01)  & 1350( 120) &  0.29(0.08) &  0.30(0.07) &  10.1(  2.4) &  10.4(  2.8)\\
  974.5 & 99.3  &  5590( 270) & 11.50(0.99)  & 1340( 115) &  0.27(0.07) &  0.27(0.06) &   9.2(  2.1) &   9.3(  2.4)\\
  976.8 & 101.6 &  5280( 140) & 11.80(0.88)  & 1350( 100) &  0.23(0.05) &  0.25(0.06) &   7.8(  1.3) &   8.6(  2.4)\\
  984.8 & 109.6 &  5660( 290) & 10.12(0.93)  & 1070(  98) &  0.22(0.06) &  0.22(0.05) &   7.5(  1.8) &   7.6(  2.0)\\
  995.5 & 120.3 &  5530( 300) &  9.80(1.42)  & 940( 136)  & 0.19(0.07)  & 0.20(0.04)  &  6.4(  2.1)  &  6.9(  1.7)\\
  999.4 & 124.3 &  5280( 290) & 10.10(0.95)  & 944(  88)  & 0.16(0.05)  & 0.18(0.04)  &  5.7(  1.4)  &  6.2(  1.6)\\
 1000.5 & 125.3 &  5550( 240) &  9.56(0.92)  & 890(  85)  & 0.18(0.05)  & 0.19(0.04)  &  6.3(  1.5)  &  6.6(  1.7)\\
 1003.9 & 128.7 &  5420( 130) & 10.60(0.80)  & 950(  71)  & 0.20(0.04)  & 0.19(0.04)  &  6.8(  1.1)  &  6.6(  1.7)\\
\hline 
\end{tabular} 
\end{minipage}
\label{lum}
\end{table*}

\begin{table*}
\centering
\caption[]{Decline rates for the total luminosity of SN 1998S} 
\begin{minipage}{\linewidth}
\begin{tabular}{ccc} \hline
Phase & Period (days) & Decline Rate (mag/day) \\ \hline
1     & 11-30   & 0.08\footnote{This rate is determined from the
blackbody-derived luminosities}\\
2     & 30-70   & 0.045\footnote{This rate is determined from the
$UVOIR$ luminosities}\\
4     & 90-130 & 0.01$^{b}$\\
\hline 
\end{tabular} 
\end{minipage}
\label{decl}
\end{table*}

The SN~1998S luminosity evolution can be divided into three phases
({\it cf.} Table~\ref{decl}):

{\bf (i) Days 11-30.}  During this phase, good fits to the photometry
are achieved with blackbody curves.  This probably reflects the fact
that the supernova spectrum is indeed close to being a blackbody at
this time.  Therefore, in spite of the large extrapolation into the
UV, it is likely that the blackbody luminosity provides a more
reliable estimate of the bolometric luminosity during this era {\it
i.e.} the $UVOIR$ luminosity severely underestimates the bolometric
luminosity.  The blackbody-derived luminosity and temperature
decreased steadily.  The temperature fell from 18000~K to 9000~K. The
photospheric radius approximately doubled in size.  To within the
uncertainties, the velocity of material {\it at} the photosphere ({\it
i.e.}  radius/epoch) remained at about 9,000~km/s.  During this
period, the emission is probably due mostly to the diffusive release
of shock-deposited energy in the expanding, cooling ejected envelope.
However, the presence of the narrow lines in the spectra (eg. Fassia
{\it et al.} 2000a) indicate that SN 1998S is surrounded by dense
circumstellar material (CSM).  Thus, reprocessing of the UV-flash by
the CSM may also contribute to the total light.
  
{\bf (ii) Days 30-70.}  After $\sim$30 days the spectrum became
increasingly affected by emission/absorption features so the blackbody
fits become poorer and less appropriate. This, together with the fact
that the proportion of the flux in the inaccessible UV is much
reduced means that the spline fits would be expected to provide a
better estimate of the bolometric luminosity in this era. Nevertheless
it can be seen from Fig.~8 that the blackbody and $UVOIR$ luminosities
steadily converge, and by about day~40 the difference is barely
significant. We believe that the use of the blackbody fits in this era
still provides useful information about the temperature, radius and
velocity.

Between days 30 and 70, the decline rate of the light curve resembles that
of type~IIL supernovae {\it viz.} 0.05~mag/day (Young \& Branch 1989).  At
first, the temperature continued to decline but then levelled off at about
5500~K by day$\sim$75.  This is close to the recombination temperature for
hydrogen.  The photospheric radius remained roughly constant at
$2-2.5\times10^{15}cm$ while the velocity of material at the photosphere
declined from $\sim$7,000~km/s to $\sim$2,500~km/s.  We can interpret the
decline in velocity as being due to the recession of the hydrogen
recombination front through the expanding ejecta, releasing shock-deposited
energy.  During this phase in the classic type~IIp event, the release of
shock-deposited energy maintains the bolometric luminosity during the
plateau phase.  However there is little sign of such a phase for SN~1998S.
We suggest that this is due to a reduced H-envelope.  Indeed, the presence
of narrow H-emission lines in the spectra (Filippenko \& Moran 1998;
Garnavich {\it et al.}  1998; Fassia {\it et al.} 2000a) indicates that the
progenitor of SN~1998S lost a considerable amount of its hydrogen envelope
prior to the explosion.  We also suggest that the steepening of the decline
at ~$\sim$70~days corresponds to the end of the plateau phase in the more
typical type~IIp event. At this epoch, the photosphere reaches the helium
mantle.  Since this has a higher recombination temperature than the
hydrogen envelope, it would have already recombined resulting in the
steeper decline of the light curve.  However, by this epoch, the
photosphere may already be encountering upwardly-mixed radioactive
$^{56}$Ni/$^{56}$Co (Fassia {\it et al.}  1998).  The locally deposited
$\gamma$-ray energy will tend to counter the recombination, thus slowing
the inward progression of the photosphere and producing the observed
decline rate of the bolometric light curve (eg. Swartz {\it et al.}  1991,
Eastman {\it et al.} 1994).

{\bf (iii) Days 100-130.} Sometime between day~80 and 100, the light
curve decline rate slowed to about 0.01~mag/day, close to the
radioactive decay rate of $^{56}$Co ({\it cf.} Fig.~\ref{lum_temp}).  By this
epoch, both the photospheric radius and the velocity of material at
the photosphere have declined sharply.  The blackbody temperature
remains at about 5500~K.  However, the increasingly nebular nature of
the spectrum means that blackbody-derived parameters are of decreasing
physical meaning.    In particular the blackbody derived
photospheric velocities for the 100-130 days era are $\sim$
1000km/s. This is substantially smaller the velocities inferred from
the wings of the emission profiles of the optically 
thin Balmer lines, $\sim$ 5000km/s (Leonard {\it et al.}
2000, Fassia {\it et al.}  2000a).  However, we note that the appearance of a broad
asymmetric and flat-top H$\alpha$ profile at day $\sim$ 100 (Leonard
{\it et al.} 2000, Fassia {\it et al.} 2000a) indicates that the
interaction of the ejecta with the circumstellar material further out
must have already began around this epoch (Leonard {\it et al.}
2000). Consequently, emission from the shocked,
high-velocity ejecta could contribute to the observed H$\alpha$ line
profile during this period (eg. Chugai \& Danziger 1994) thus 
increasing its width. 

The decline rate of the light curve during this period exhibits a
remarkably good match to the radioactive decay luminosity decline rate
of $^{56}$Co ({\it cf.} Fig.~\ref{lum_temp}).  We therefore suggest
that, by this era, all the shock-deposited energy has been released and
the ejecta are being powered by radioactive decay energy.  Moreover,
{\it all} the $\gamma$-rays and positrons are still being trapped in
the ejecta at this phase.  We can therefore estimate the total mass of 
$^{56}$Ni produced in the explosion. On the basis of the $UVOIR$
luminocities on days 109.6--128.7, and assuming an explosion date of
1998 February~28.0, we find a total initial $^{56}$Ni mass of
0.15$\pm$0.05 M$_{\odot}$.  Of course, as mentioned above, the H$\alpha$
line profile during this epoch indicates that interaction of the ejecta
with the CSM must have started. Thus emission from the interaction
could be adding to the bolometric luminosity.  However, given the match
to the $^{56}$Co radioactive decline rate we conclude that the
proportion of this contribution is less than 15\%.  It should be
remembered that the derived mass of $^{56}$Ni is subject to any
uncorrected systematic errors in the extinction correction (see
Section~3).

\subsection{Comparison with SN 1988Z}
In Figure~\ref{98Soplc}   we showed the $B-V$ evolution of the well-observed type~IIn
SN~1988Z (Turatto {\it et al.} 1993).  Clearly, its behaviour was
rather different from that of SN~1998S.  During its first 100~days, SN
1988Z declined more slowly ({\it cf.} Figure~\ref{98Sbvri}).  Turatto
{\it et al.} (1993) suggested that the slowly declining light curve of
SN 1988Z indicated the ejection of a massive, extended envelope in the
explosion.  However, Chugai \& Danziger (1994) modelled the evolution
of the H$\alpha$ profile of SN 1988Z and concluded that the ejected
envelope was $<$ 1 M$_{\odot}$. They argued that SN 1988Z originated
from a 8-10 M$_{\odot}$ main sequence star which lost most of its mass
through a high mass-flow wind which lasted right up to the explosion.
Chugai (1997) estimates a mass loss rate of
$\dot{M}\sim7\times10^{-4}u_{10}M_{\odot}/yr$ where $u_{10}$ is the
wind velocity in units of 10km/s.  In this scenario the interaction of
the ejecta with the CSM began very soon after the explosion.  This
interaction produced the observed high luminosity and slow-declining
light curve, masking the thermal emission of the supernova ejecta
(Chugai \& Danziger 1994, Chugai 1997).

The optical light curves of SN 1998S decline faster than those of
SN~1988Z and, as indicated in the previous section, resemble more the
light curve of the linear type II supernova (e.g. SN 1979C, Barbon {\it
et al.} 1982). Linear type II supernovae are thought to arise from
stars with main sequence masses $\sim$ 8-20 M$_{\odot}$ (Blinnikov \&
Bartunov 1993). Similar to the scenario suggested for SN 1988Z by
Chugai \& Danziger (1994), these stars lose most of their envelope mass
through stellar winds prior to explosion.  However, the winds are
considerably weaker. For SN 1979C, Chugai (1997) estimated a rate of
$\dot{M}\sim10^{-4}u_{10}M_{\odot}/yr$, yielding an envelope of $\sim$2
M$_{\odot}$ at the time of the explosion.  We therefore suggest that
the fast decline of the optical light curves of SN~1998S indicates that
the mass of the supernova-ejected envelope was quite low but that the
wind was weaker than in the case of SN~1988Z.  Thus, at least in the
first few months, the ejecta expanded without significant interaction.

Based on optical spectra of SN 1998S, Leonard et al. (2000) suggest
that the progenitor of SN 1998S underwent a significant mass-loss
episode that ended $\sim$60 years prior to the explosion. They also
suggest that a second, weaker, mass-loss episode began seven years
prior to the explosion. When the star exploded it immediately
interacted with the most recent wind completely engulfing it within a
few days. The interaction with the older, more massive wind started
after $\sim$100 days.

\subsection{$H-K$ and $K-L'$ excess} 
The $H-K$ colour around day~+17 is about +0.03 which is consistent with
the (reddened) blackbody temperature given in Table~4.  However, by
day~+76.7, the $H-K$ colour exhibits an excess of at least 0.15 mag
relative to the blackbody temperature.  This excess continued to
increase attaining $\sim0.6$ by day~128.7.  Up to day 110, the
appearance and growth of this excess is due mostly to CO
emission. $K$-band spectra shown in Fassia {\it et al.} (2000a) on
day~109, and in Gerardy {\it et al.}  (2000) on day~110 (using our
definition of epoch 0~days) show strong emission in the CO first
overtone band.  Beyond day 110, however, thermal emission from dust
(see below) may have made an increasingly significant contribution to
the $H-K$.  As we argue below, cool dust ($T\le$800K) is present in the
vicinity of the supernova. Thermal emission from such cool dust would
increase the $K$ band flux whereas it would have a negligible
contribution to the $H$ band flux.

\begin{figure}
\vspace{9cm}
\includegraphics{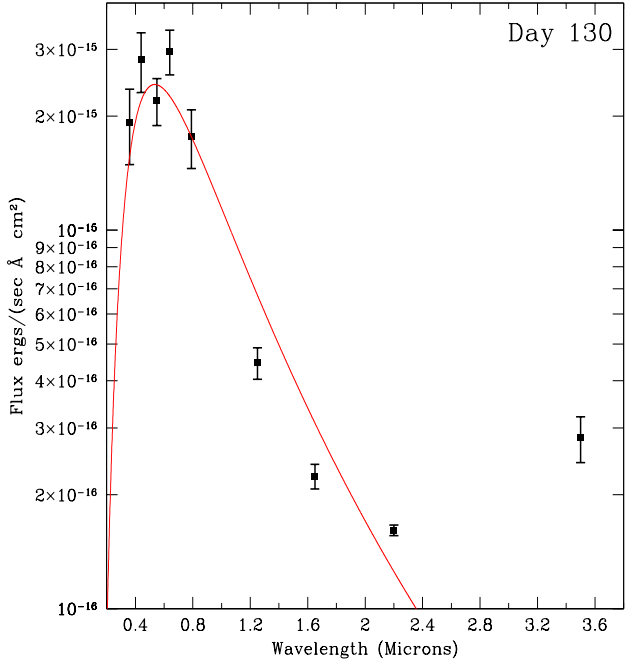}
\caption[]{Blackbody fit to the photometric data obtained on day
130. The excess in the infrared is clearly evident. }
\label{irex}
\end{figure}

On day~130 we observed $K-L'=2.55\pm0.15$ mag.  Yet, for the
temperature deduced from the blackbody fit on day~130 ($\sim$6000~K)
we would expect a $K-L'$ colour of only about +0.15 (including
extinction reddening).  This is illustrated in Fig.~\ref{irex}.
Moreover, we know that there was strong first overtone CO emission in
the $K$-band at about this time, which would tend to produce a $K-L'$
colour even bluer than +0.15.  We are unaware of any species which
would produce sufficient line emission in the $L'$-band to account for
the very red $K-L'$ colour.  We also know that by day~+324 IR spectra
(Fassia {\it et al.} 2000b) show a $K$-band {\it continuum} which
rises steeply towards long wavelengths.  We believe therefore that the
extremely red $K-L'$ indicates the presence of an additional continuum
source of IR emission.

The occurrence of such an IR excess has been observed in other supernovae
({\it e.g.} Graham {\it et al.} 1984; Dwek 1983) and is commonly
attributed to emission from warm dust.  We believe that this is also the
case here.  We wish to determine the origin of the dust and its energy
source.  The flux of thermal emission from dust at a given temperature
must come from a region at least as large as a blackbody of the same
temperature.  Constraints on the epoch of dust condensation can therefore
be obtained by comparing the flux from the dust on day~130 with the size
of the supernova at the same epoch.  For a dust temperature of 1500~K
({\it i.e.} close to evaporation temperature) and a distance of
$\sim$17~Mpc to NGC~3877, the $L'$ magnitude (+11.1) on day~130 requires a
blackbody radius of at least $\sim$ 10$^{16}$ cm. If dust formed in the
ejecta after the explosion, to achieve this radius by day~130 the dust (or
its constituent elements and molecules) would have to travel with a
velocity of $\sim$10700~km/s.  Such a high velocity would be associated
with the hydrogen envelope.  However, densities and abundances favourable
to dust condensation are only likely to occur much deeper in the ejecta,
in the slow-moving layers of metals (Graham {\it et al.} 1983).  Moreover,
the spectra of SN~1998S show no evidence of fast-moving metals (Fassia
{\it et al.} 2000a).  Thus we conclude that most of the thermal emission
from dust observed on day~130 originates from grains which condensed prior
to SN~1998S, in the stellar wind of the progenitor. 
On days~325 and ~355 respectively, Fassia {\it et al.} (2000b) and
Gerardy {\it et al.}  (2000) observed a near-IR continuum rising
towards longer wavelengths.  Gerardy {\it et al.} argue that the
observed continuum is likely due to dust heated by the CSM-ejecta
interaction either by direct shock-heating or by absorption of X-rays
from the interacting region. However, they are unable to distinguish
whether the emitting dust is pre-existing dust in the CSM or newly
formed dust in the supernova ejecta. We note that the attribution of
the strong IR-excess, observed on day~130, to pre-existing dust in the
CSM does not contradict the dust condensation scenario. Dust could have
formed in the ejecta at later epochs.  The observed $K-L'=+2.5$~mag ,
corresponds to a blackbody temperature of $\sim800$~K.  However,
spectral energy distributions ($JHKLM$) measured by us (Fassia {\it et
al.} 2000b) between 130 and 691~days post explosion revealed that by
324~days, the temperature had risen to nearly 1300~K. It subsequently
fell to $\sim900$K by 691~d.  This later, hotter IR emission could be
due to an episode of dust condensation in the ejecta.  We shall
investigate this further in Fassia {\it et al.} (2000b).

How might circumstellar dust be heated at epoch 130~days?  One
possibility is that optical/UV photons emitted from the supernova
around maximum light heat the dust which then re-radiates this energy
in the thermal IR.  Light-travel time effects cause a delay in the
reception of the IR flux at earth.  This is usually referred to as an
``IR-echo'' (Bode \& Evans 1980; Dwek 1983; Graham {\it et al.}  1984;
Graham \& Meikle 1986; Roche {\it et al.} 1989; Felten \& Dwek 1989).
Alternatively, the dust could be heated by X-rays from the ejecta-CSM
shock front.  At epoch 130~days, given the data available, it is
difficult to choose between these two scenarios.  However, second- and
third-season observations show a strong $K-L'$ excess persisting to at
least day~690.  The energy required to maintain the IR emission for
such a long time probably favours the latter (X-ray) scenario, at
least for later times.  This will be discussed further in Fassia {\it et
al.} (2000b).

\section{Summary} 
We have presented infrared and optical photometric observations of SN~1998S
covering the first $\sim$130 days after explosion.  This is the first time
that such extensive wavelength coverage has been achieved for a type~IIn
event.  A value of $A_{V}=0.68\pm0.06$ mag is derived from interstellar
Na~I~D lines measurements. Using our photometry we carried out blackbody
and spline fits to determine the bolometric light curve.  During the first 2--3
months, the luminosity is dominated by the the release of shock-deposited
energy in the ejecta.  During the first month, the photosphere stayed at
about the same position with respect to the ejecta and the shock-deposited
energy escaped by diffusion up to the photosphere.  Subsequently, the shock
energy was released as the photosphere receded inwards through the hydrogen
envelope. After day~70, the bolometric light curve exhibited a steepening
decline rate which we associate with the recombination front reaching the
hydrogen-helium boundary. By day~100 the light curve decline rate slowed to
about 0.01~mag/day matching the radioactive decay rate of $^{56}$Co.  This
agreement indicates that, by this time, the luminosity was dominated by the
deposition of radioactive-decay energy and that all the $\gamma$-rays and
positrons were trapped in the ejecta.  From the bolometric luminosity after
day$\sim$110 we estimate that 0.15$\pm$0.05 M$_{\odot}$ of $^{56}$Ni was
produced in the explosion.  On day~130 a strong IR ``excess'' was observed.
We have argued that this must be due to {\it pre-existing} dust in the
circumstellar material of SN 1998S.  The dust could be heated either by the
initial UV/optical flash of the supernova or by the X-rays from the
CSM-shock front interaction.

\section*{Acknowledgements} 
We would like to thank Diane Harmer for co-ordinating the WIYN observations
and Paul Smith and A. Saha for obtaining some of the WIYN observations. We
are also very grateful to Bob Carswell who provided us with the software
that fitted Voigt profiles to spectroscopic data. S. Bennett is supported
by a PPARC studentship. AAS acknowledges generous financial support from
the Royal Society. The JKT is operated on the island of La Palma by the
Isaac Newton Group in the Spanish Observatorio del Roque de los Muchachos
of the Instituto de Astrofisica de Canarias. The WIYN Observatory is a
joint facility of the University of Wisconsin-Madison, Indiana University,
Yale University, and the National Optical Astronomy Observatories.  The
IAC80 telescope is at the Observatorio del Teide, operated by the Instituto
de Astrof'isica de Canarias. Finally we acknowledge that some of the data
reported here were obtained as part of the UKIRT Service Programme.

\end{document}